\documentclass{article} %
\usepackage[preprint]{colm2025_conference}

\usepackage{graphicx}

\usepackage{microtype}
\usepackage{hyperref}
\usepackage{url}
\usepackage{booktabs}
\usepackage{multirow}
\usepackage{colortbl}
\definecolor{LightCyan}{rgb}{0.75,1,1}

\usepackage{lineno}

\definecolor{darkblue}{rgb}{0, 0, 0.5}
\hypersetup{colorlinks=true, citecolor=darkblue, linkcolor=darkblue, urlcolor=darkblue}

\usepackage{xspace}
\usepackage{todonotes}

\usepackage{wrapfig}

\usepackage{pifont}%

\title{Rank-K: Test-Time Reasoning for Listwise Reranking}

\author{Eugene Yang
\quad Andrew Yates 
\quad Kathryn Ricci 
\quad Orion Weller\\
\textbf{Vivek Chari}
\quad \textbf{Benjamin Van Durme}
\quad \textbf{Dawn Lawrie}
\\
  Johns Hopkins University\quad 
  \\
  \texttt{\small{\{eugene.yang, andrew.yates, vandurme\}@jhu.edu}}
}

\begin{document}
\ifcolmsubmission
\linenumbers
\fi

\maketitle

\begin{abstract}

Retrieve-and-rerank is a popular retrieval pipeline because of its ability to make slow but effective rerankers efficient enough at query time by reducing the number of comparisons. 
Recent works in neural rerankers take advantage of large language models for their capability in reasoning between queries and passages and have achieved state-of-the-art retrieval effectiveness. 
However, such rerankers are resource-intensive, even after heavy optimization. 
In this work, we introduce Rank-K, a listwise passage reranking model that leverages the reasoning capability of the reasoning language model at query time that provides test time scalability to serve hard queries. 
We show that Rank-K improves retrieval effectiveness by 23\% over the RankZephyr, the state-of-the-art listwise reranker, when reranking a BM25 initial ranked list and 19\% when reranking strong retrieval results by SPLADE-v3.
Since Rank-K is inherently a multilingual model, we found that it ranks passages based on queries in different languages as effectively as it does in monolingual retrieval.
\end{abstract}

\section{Introduction}

Neural retrieval models have benefited from advancements in language modeling and have achieved significant improvements in retrieval effectiveness~\citep{DL19, DL20}. 
While much of the work has focused on encoding the queries and documents separately to enable offline document processing~\citep{wang2022text, dpr, colbert, yangecir24, formal2021splade,weller2024promptriever}, processing queries and documents jointly, while inefficient, has been pushing the state-of-the-art in retrieval effectiveness for the last several years~\citep{colbertv2, lassance2024splade, rocketqa}. 
These models, such as MonoBERT~\citep{monobert} and MonoT5~\citep{monot5}, rerank the top-ranked documents from another upstream retrieval model to reduce query latency by limiting the amount of computation. 

As the reasoning capability of language models has improved, using a generative model as a reranker has become a widely used component in designing search systems~\citep{reddy2024first, sun2023rankgpt,pradeep2023rankvicuna}. 
For high-stakes use cases where users will tolerate higher latency, attaching a less efficient but very effective model at the end of the pipeline can greatly improve the overall utility of the system~\citep{teevan2014slow}. 
In this work, we explore using generative models with built-in reasoning capabilities to rerank. 

Pointwise reranking produces a score for a query-passage pair by jointly encoding them through a language model~\citep{monot5, monobert}. The ranking is then inferred by sorting the scores of the query-passage pairs. 
While architecturally easier to train since it is identical to a sequence pair classifier, a pointwise reranker does not compare different passages directly when producing the final ranking. 

Alternatively, a listwise reranker~\citep{sun2023rankgpt, pradeep2023rankvicuna, pradeep2023rankzephyr} compares a set of passages to a query to directly produce a ranked list based on the query. 
Instead of comparing passages through scores, listwise rerankers compare the content of the passages before producing the final ranking. 
Prior work in using generative models as listwise rerankers only prompt the generative model to produce the ranking without reasoning~\citep{pradeep2023rankvicuna, pradeep2023rankzephyr}. 
To leverage the recent advancements in reasoning models, we ask the generative model to provide a reasoning trace that analyzes and compares passages to produce 
the ranking. 

\begin{figure}
    \centering
    \includegraphics[width=\linewidth]{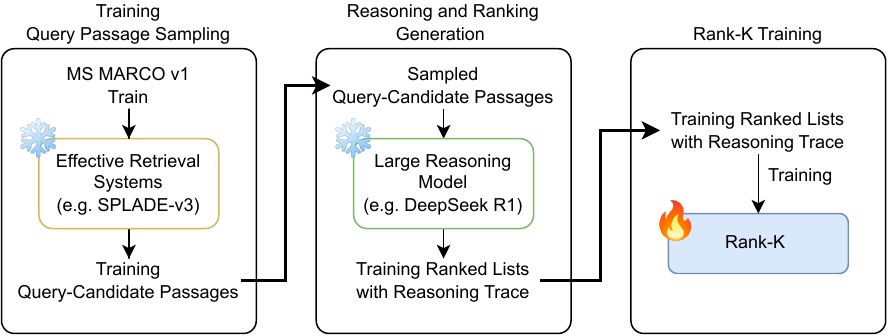}
    \caption{Overview of the training pipeline for Rank-K. }
    \label{fig:rankk-overview}
\end{figure}

We introduce Rank-K, a listwise passage reranking model with test-time reasoning ability to produce a more accurate ranking that is fine-tuned from QwQ 32B with distillation from the reasoning trace produced by DeepSeek R1 with explicit ranking instruction.
An overview of our training pipeline is summarized in Figure~\ref{fig:rankk-overview}.
The training reasoning traces, Rank-K model, and supporting scripts are available on GitHub and HuggingFace.\footnote{\url{https://github.com/hltcoe/rank-k}}
We show the Rank-K is more effective than the current state-of-the-art listwise reranker in reranking English passages. 
Since Rank-K is naturally multilingual, it also demonstrates a strong cross-language ability in ranking queries and passages in different languages through zero-shot language transfer. Thus, our work contributes the following by:

\begin{itemize}
    \item Introducing the first listwise reranking model that utilizes test-time reasoning, 
    \item Releasing a training collection with accurate reranking and reasoning traces for distillation, and 
    \item Providing evaluation results demonstrating the effectiveness of Rank-K in both English and cross-language reranking. 
\end{itemize}

\section{Background}

Modern information retrieval pipelines balance effectiveness and efficiency by factoring retrieval into multiple stages of computation at query time~\citep{lin2022pretrained}. In the first stage, an efficient retriever, such as BM25~\citep{robertson2009probabilistic}, E5~\citep{wang2022text}, and PLAID~\citep{yangecir24, plaid}, scores pre-computed (sparse or dense) vector representations of all documents against the representation of the query. In the subsequent stages, rerankers that are more effective but computationally more expensive, because of their inability to preprocess the documents offline, are used to improve the order of a subset of documents to limit the latency. Among them, cross-encoders, such as MonoBERT~\citep{monobert} and MonoT5~\citep{monot5}, leverage cross-attention to compute a query-aware final ranking of the top-k documents from the upstream stages. 
This pipeline approach gradually funnels a large collection down to a handful of documents~\citep{lassancetrec23}. 
While some highly effective rerankers that use a large model underneath are slow, limiting the number of processing documents makes them practical in a retrieval system. 
In this work, we focus on evaluating reranker effectiveness.

Analogous to early learning-to-rank work, there are primarily three types of rerankers: pointwise, pairwise, and listwise models~\citep{cao2007learning}, which process the query with a document, a pair of documents, and a list of documents, respectively.
When chaining multiple rerankers together in a pipeline, the number of passages (the \textit{depth}) processed by each component usually decreases~\cite{lin2022pretrained}.
For example, a pointwise reranker might rerank the top-100 documents, followed by an expensive pairwise reranker that produces a final ranking by considering all pairs among the top-20 documents~\citep{pradeep2021expando, lin2022pretrained}.
While there are efficiency tricks to reduce the amount of computation without sacrificing the reranking depth, such as quantization~\citep{jacob2017antization} and better batching~\citep{sun2023rankgpt, zhuang2024setwise}, they are orthogonal to this work. As such, we exclude these efficiency tricks and focus on improving the core modeling ability.

Recent advancements in reinforcement learning have greatly enhanced the reasoning capabilities of LLMs, such as DeepSeek R1~\citep{deepseekai2025deepseekr1}, QwQ~\citep{qwq32b}, and o1~\citep{jaech2024openai}. Such models are endowed with an improved ability to perform chain-of-thought \citep{wei2022chain} reasoning, enabling them to analyze complex user inputs and systematically evaluate multiple pieces of evidence. While prior work in listwise re-ranking has focused on directly supervising the output of the re-ranker~\citep{sun2023rankgpt, pradeep2023rankvicuna}, reasoning capabilities offer an opportunity to further improve listwise reranking. In particular, using a chain-of-thought process to generate a ranking encourages logical decision-making and allows for interpretable rankings. Additionally, recent work has suggested that, under appropriate fine-tuning, small models can achieve reasoning capabilities that are competitive with much larger models~\citep{shridhar2023distilling, weller2024followir,weller2025rank1}, suggesting the feasibility of training reasoning rerankers efficiently enough to be integrated into a retrieval pipeline.

\begin{figure*}
\noindent\fbox{%
    \parbox{.98\textwidth}{%
\footnotesize
{\tt
Determine a ranking of the passages based on how relevant they are to the query. 
If the query is a question, how relevant a passage is depends on how well it answers the question. 
If not, try analyze the intent of the query and assess how well each passage satisfy the intent. 
The query may have typos and passages may contain contradicting information. 
However, we do not get into fact-checking. We just rank the passages based on they relevancy to the query. 

Sort them from the most relevant to the least. 
Answer with the passage number using a format of `[3] > [2] > [4] = [1] > [5]`. 
Ties are acceptable if they are equally relevant. 
I need you to be accurate but overthinking it is unnecessary.
Output only the ordering without any other text.

Query: \{query\} 

[1] \{passage 1 content\}

[2] \{passage 2 content\}

...
}
}}

\caption{Prompt for Reasoning and Ranking Passages}
\label{prompt}
\end{figure*}

\section{Rank-K Reranking}

In this section, we discuss the process of training Rank-K, as illustrated in Figure~\ref{fig:rankk-overview}.

\subsection{Training Data Sampling}

Typically, when training a pointwise ranker, we need to gather supervised learning signals for query-passage pairs. 
However, for listwise rankers, the supervised signal we need is a complete ranking of a set of passages for a given query. 
We construct our training set based on queries and passages from MS MARCO v1 passage collection~\citep{nguyen2016ms}, which has been shown to be effective in training retrieval models. 

We sample 50k queries, which is roughly 10\% of the training queries in the collection after deduplication. 
For each query, we construct a ranked list of passages combining three retrieval pipelines using reciprocal rank fusion~\citep{cormack2009reciprocal}: PLAID-X~\citep{yang2024distillation} followed by Mono-mT5 reranking~\citep{jeronymo2023neuralmind}, CoCondenser~\citep{gao2022unsupervised}, and SPLADE-v3~\citep{lassance2024splade}.
To ensure a diversity of difficulty in the training data, we create two passage sets by taking passages from ranks 1 to 20 and from ranks 10 to 30. 
Each sample contains 10 to 20 passages, for which the length is uniformly sampled. 
This sampling process results in a set of 100k queries-passage set pairs, with the number of passages varying from 10 to 20. 

\subsection{Example Generation}

To generate the rankings for supervision, we prompt DeepSeek R1~\citep{deepseekai2025deepseekr1} with specific directions to reason and rank the passages based on their relevance to the associated query. 
The structure of the prompt is similar to prior work, such as RankGPT~\citep{sun2023rankgpt} and RankZephyr~\citep{pradeep2023rankzephyr}, where the content of the passage is attached with a numeric ID in the context. The output is a list of numeric IDs, indicating the ranking. Our prompt is shown in Figure~\ref{prompt}.

We designed the instruction to mimic how a human would have assessed the passages' relevance. 
First, the passages can be tied, which prevents the model from overthinking the ordering between two passages that are near duplicates or identical.
Second, since DeepSeek R1 is a reasoning model, the thinking tokens also need to be generated, where a longer thinking process results in higher query latency and computational cost. We instruct the model not to overthink the problem to prevent the model from generating too many thinking tokens and exhaust the context window before producing the final ranking. 
Third, we provide a more detailed definition of relevance. Unlike RankZephyr, which simply asks the model to rank passages based on \textit{relevance}, if the query is indeed a question, we instruct the model to judge relevance based on the ability of the passage to answer the question. If not, we instruct the model to investigate the intent of the query and how well the passage satisfies it.
This provides more guidance and robustness to the model for the general ranking task. 
Finally, we explicitly ask the model not to fact-check the provided passages using parametric knowledge. 
These instructions ensure that the model produces an accurate ranking with a reasonable output length. We iterated on the prompt using TREC Deep Learning 2022~\citep{DL22} as a development data set. This data set was not used in any of experiments.

\begin{wraptable}{r}{7cm}
\centering
\begin{tabular}{l|rr}
\toprule
           & \textbf{Input}  & \textbf{Output}  \\
\midrule
Per Call    &   1,024.65  &    2,408.85 \\   
Per Passage &      68.76  &      166.28 \\
\bottomrule
\end{tabular}
    \caption{Average number of tokens in input and output per call and per passage for generating the training data using DeepSeek R1. }
    \label{tab:training-data-stats}
\end{wraptable}

Table~\ref{tab:training-data-stats} summarizes the statistics of the generated data. 
The amount of generated tokens, including the thinking tokens, is 2.4 times the number of input tokens on average, with a 90\% percentile of 3,957 output tokens. 
From investigating a handful of examples from the generated data, the reasoning trace usually first summarizes each document with several phrases, categorizes them into buckets, produces partial rankings among each bucket, assembles the full ranking, and verifies the result through self-reflection. 
This long process requires a large number of generated tokens to support the reasoning. 

We leverage Kluster.ai to generate these data in batch. The process took around 36 hours and around USD\$1,900 to complete using a temperature of 0.7.

\subsection{Reasoning Distillation}

Finally, we use the generated examples to train a smaller reasoning language model with LoRA for the listwise reasoning and reranking task.
The smaller language model receives the same input as DeepSeek R1 and is fine-tuned to generate the same output that R1 generated.
Specifically, we train QwQ 32B~\citep{qwq32b}, which is a reasoning model extended from Qwen2.5 32B~\citep{yang2024qwen2} that achieves similar reasoning ability to DeepSeek R1 and o1-mini~\citep{jaech2024openai}.
We use 128 AMD MI250X GPUs to train the QwQ 32B model with LoRA using LLaMA-Factory~\citep{zheng-etal-2024-llamafactory}. We train for two epochs with a batch size of 128 and a learning rate of 1e-4.
We call the resulting model the Rank-K model. 

To investigate the impact of the reasoning tokens, we also train QwQ 32B model using only the final ranking from R1.
To do so, we extract only the final ranking from the generated data and train QwQ 32B for two epochs as described above.
We call this model \textit{Naive} Rank-K since it does not reason about the reranking process at test time. 

\section{Experiments}

We test Rank-K on three retrieval datasets, TREC Deep Learning (DL) track 2019 and 2020~\citep{DL19, DL20}, TREC NeuCLIR track 2022 to 2024~\citep{neuclir2022, neuclir2023}, and BRIGHT~\citep{bright}.\footnote{Since the BRIGHT collection is constantly changing after their first release, we use the version that MTEB~\citet{muennighoff2022mteb,enevoldsen2025mmteb} uses, which is at commit \texttt{a75a0eb} in the BRIGHT Huggingface repository (\url{https://huggingface.co/datasets/xlangai/BRIGHT/tree/a75a0eb}). }
Since NeuCLIR is a collection for cross-language retrieval, where queries are in English and documents are in Chinese (Zho), Persian (Fas), and Russian (Rus), we use the translated documents, which are in English, that are available publicly on Huggingface Datasets.\footnote{\url{https://huggingface.co/datasets/neuclir/neuclir1}} 
The summary statistics of the two TREC evaluation datasets are summarized in Table~\ref{tab:collection}.
Please refer to the BRIGHT dataset paper for the full statistics of the dataset. 
For NeuCLIR, we use the title and description fields from the topics as the queries. 
Given the NeuCLIR and BRIGHT documents can be long, all reranking experiments are run by truncating documents at 450 tokens.

\begin{table}[t]
    \centering
\begin{tabular}{ll|cc|c}
\toprule
&    & \# of Queries & \# Docs & Avg. Doc Length \\
\midrule
\multirow{2}{*}{Deep Learning}
& 2019  &       43   & \multirow{2}{*}{8,841,823} & \multirow{2}{*}{334.79}  \\
& 2020  &       54   &         & \\
\midrule
\multirow{3}{*}{NeuCLIR}
& Persian &    174   &  2,232,016 &  399.37 \\
& Russian &    171   &  4,627,541 &  307.91\\
& Chinese &    167   &  3,179,206 &  372.52\\

\bottomrule
\end{tabular}
    \caption{Collection Statistics. Document length is the number of estimated tokens separated by whitespaces.}
    \label{tab:collection}
\end{table}

To conduct a fair comparison between rerankers, we hold the first stage retrieval result constant. 
In this work, we rerank the top 20 documents from BM25 and SPLADE-v3 for TREC DL and TREC NeuCLIR. 
Unlike prior work, we report reranking documents without grouping candidates into batches to support reranking more than a model call can take at one time. 
Since these approaches essentially work as a bubble sort (or a partial bubble sort) to cope with the context length limitation~\citep{sun2023rankgpt, pradeep2023rankzephyr} by using a sliding window to rerank a subset of the documents, we argue that the document batching is orthogonal to the modeling and effectiveness of the reranker. 
Therefore, we primarily report nDCG@10, which only looks at the top 10 documents after reranking.
For BRIGHT, we report reranking BM25 and ReasonIR that takes both the original query and the GPT-4o reasoning trace, while the rerankers only use the original queries. 

To provide compatibility with other recent and concurrent evaluated on BRIGHT, we additionally conduct an ablation study on 
using a single-iteration sliding window approach from the bottom of the initial rank list. We rerank the top 100 documents with a window of 20 
(identical to our main results) and a stride of 10. Therefore, each query requires 10 Rank-K reranking calls. 

As a baseline, we report results with RankZephyr, a state-of-the-art listwise reranker model fine-tuned from Mistral-7B~\citep{jiang2023mistral}, and Rank1~\citep{weller2025rank1}, a pointwise reranker that also leverages the reasoning ability of DeepSeek R1 that comes with 7B and 32B versions. %
Both RankZephyr and Rank1 are run using RankLLM\footnote{\url{https://github.com/castorini/rank_llm}}, a toolkit for orchestrating LLM rerankers. 
These two models represent two critical comparison points that Rank-K shares similarities in model architecture with. 
Additionally, we compare the reranking results with the first stage retrieval results, over which, surprisingly, some rerankers fail to improve. 

\section{Results and Analysis}

Table~\ref{tab:main-results} summarizes the reranking experiment results.
The Rank-K model provides a significant improvement over RankZephyr, which is the previous state-of-the-art listwise reranking model. 
While improvements are small on the TREC DL 2019 and 2020 benchmarks, which have been heavily exposed in hyperparameter tuning~\citep{parry2025variations}, evaluating on the translated TREC NeuCLIR data demonstrates a strong robustness of the Rank-K model over RankZephyr when moving beyond the training collection (both Rank-K and RankZephyr were trained on MS MARCO v1 passages though they differ in sampling and supervised signal generation). 

Particularly, RankZephyr failed to improve the first-stage ranking on TREC NeuCLIR, resulting in a macro-average only 2\% improvement and even 6\% \textit{degradation} in nDCG@10 over the first-stage ranking produced by either BM25 and SPLADE-v3, respectively. 
Such differences in reranking BM25 and SPLADE-v3 signifies the importance of evaluating with a strong first-stage retriever, both because it is a more challenging task to distinguish passages that are all likely to be relevant and because reranking a strong retriever like SPLADE-v3 is a more realistic setting for users.

While both Rank-K and Rank1 are reasoning rerankers with similar number of parameters, the pointwise Rank1 is significantly worse than the listwise Rank-K on both DL and NeuCLIR. 
The differences are consistent across the two collections and reranking the two first stage retrievers, showing a strong effectiveness gain for the model to reason across passages at test time. 
However, similar to Rank-K, Rank1 also demonstrates strong robustness in reranking passages beyond MS MARCO.
We suspect that the reasoning process provides a strong robustness when transferring to other domains.

Comparing Rank-K with its teacher model, DeepSeek R1, the teacher model is slightly more effective than Rank-K when reranking BM25 but, on average, there is no difference when reranking a strong SPLADE-v3 initial ranking. 
However, because of the model size (32B vs 671B), Rank-K is much more efficient than simply prompting R1. 
While there is some knowledge that was not distilled successfully from the teacher to Rank-K, such a knowledge gap does not seem to affect the ability to compare passages where the differences are more nuanced, based on the near-identical effectiveness in reranking SPLADE-v3.

\begin{table}
\renewcommand{\b}[1]{\textbf{#1}}
\centering

\begin{tabular}{ll|cc|ccc|p{15.5pt}c}
\toprule
First Stage &  \multirow{2}{*}{Reranker}          & \multicolumn{2}{c|}{TREC DL} & \multicolumn{3}{c|}{TREC NeuCLIR}  & \multicolumn{2}{c}{\multirow{2}{*}{Avg. \, ($\Delta$)}} \\
Retriever &   &  2019 &  2020 & Fas. & Rus. & Zho. & & \\
\midrule

\multirow{7.5}{*}{BM25}
& --         & 0.499 & 0.485 & 0.386 & 0.337 & 0.375 & 0.416 &   \\
& RankZephyr & 0.645 & 0.633 & 0.291 & 0.279 & 0.274 & 0.424 &  $(2\%)$ \\
& Rank1-32B  & 0.649 & 0.612 & 0.420 & 0.419 & 0.432 & 0.506 &  $(22\%)$ \\
& R1 Teacher & 0.657 & 0.629 & 0.482 & 0.474 & 0.470 & 0.543 &  $(30\%)$ \\
\cmidrule{2-9}
& {\cellcolor{LightCyan}}Naive Rank-K & {\cellcolor{LightCyan}}0.617 & {\cellcolor{LightCyan}}0.578 & {\cellcolor{LightCyan}}0.430 & {\cellcolor{LightCyan}}0.411 & {\cellcolor{LightCyan}}0.425 & {\cellcolor{LightCyan}}0.492 & {\cellcolor{LightCyan}}$(18\%)$ \\
& {\cellcolor{LightCyan}}Rank-K & {\cellcolor{LightCyan}}0.662 & {\cellcolor{LightCyan}}0.643 & {\cellcolor{LightCyan}}0.440 & {\cellcolor{LightCyan}}0.434 & {\cellcolor{LightCyan}}0.447 & {\cellcolor{LightCyan}}0.525 &{\cellcolor{LightCyan}}  $(26\%)$ \\
\midrule
\multirow{7.5}{*}{SPLADE-v3}
& --         & 0.722 & 0.751 & 0.507 & 0.449 & 0.428 & 0.571 &   \\
& RankZephyr & 0.775 & 0.804 & 0.398 & 0.406 & 0.314 & 0.539 &  (-6\%) \\
& Rank1-32B  & 0.762 & 0.762 & 0.545 & 0.496 & 0.482 & 0.609 &  $(7\%)$\\
& R1 Teacher & 0.767 & 0.786 & 0.601 & 0.535 & 0.534 & 0.645 &  $(13\%)$ \\
\cmidrule{2-9}
 & {\cellcolor{LightCyan}}Naive Rank-K & {\cellcolor{LightCyan}}0.757 & {\cellcolor{LightCyan}}0.775 & {\cellcolor{LightCyan}}0.552 & {\cellcolor{LightCyan}}0.516 & {\cellcolor{LightCyan}}0.504 & {\cellcolor{LightCyan}}0.621 &  {\cellcolor{LightCyan}}$(9\%)$ \\
& {\cellcolor{LightCyan}}Rank-K  & {\cellcolor{LightCyan}}0.785 & {\cellcolor{LightCyan}}0.799 & {\cellcolor{LightCyan}}0.581 & {\cellcolor{LightCyan}}0.528 & {\cellcolor{LightCyan}}0.518 & {\cellcolor{LightCyan}}0.642 &  {\cellcolor{LightCyan}}$(12\%)$ \\

\bottomrule
\end{tabular}
\caption{nDCG@10 on TREC Deep Learning and NeuCLIR tracks. All rerankers rerank the top 20 passages from the first stage retriever. Our contribution is highlighted in blue.}
\label{tab:main-results}
\end{table}

By comparing Rank-K with Naive Rank-K, which excluded the reasoning trace from fine-tuning, the difference in effectiveness shows that reasoning trace provides a strong performance improvement. 
Despite the degradation over Rank-K, Naive Rank-K is still significantly better than RankZephyr, which is likely to be a result of better training example sampling (since RankZephyr also uses MS MARCO v1 as the corpus for distillation) and a more effective teacher model.

\subsection{BRIGHT Evaluation}

\begin{table*}[t!]
\centering
\resizebox{\textwidth}{!}{
\begin{tabular}{l|rrrrrrr|rr|rrr|r}
\toprule
& \multicolumn{7}{c|}{StackExchange} & \multicolumn{2}{c|}{Coding} & \multicolumn{3}{c|}{Theorem-based} & \multirow{2}{*}{\centering Avg.}\\
\cmidrule(r){2-8} \cmidrule(r){9-10} \cmidrule(r){11-13}
& Bio. & Earth. & Econ. & Psy. & Rob. & Stack. & Sus. & Leet. & Pony & AoPS & TQ. & TT. \\
\midrule
BM25  & 19.2 & 27.1 & 14.9 & 12.5 & 13.5 & 16.5 & 15.2 & 24.4 & 7.9 & 6.0 & 13.0 & 6.9 & 14.8\\
BM25 on GPT-4o CoT & 53.6 & 53.6 & 24.3 & 38.6 & 18.8 & 22.7 & 25.9 & 19.3 & 17.7 & 3.9 & 18.9 & 20.2 & 26.5  \\
ReasonIR on GPT-4o CoT &  43.5 & 43.0 & 32.7 & 38.9 & 20.8 & 30.9 & 27.3 & 31.5 & 19.7 & 7.4 & 35.0 & 36.6 & 30.6 \\
\midrule
\multicolumn{14}{l}{\textit{Rerank Top 20}} \\
\midrule
RankZephyr 7B & 44.2 & 30.5 & 17.9 & 28.2 & 17.7 & 27.0 & 21.9 & 18.9 & 17.7 & 2.5 & 16.0 & 11.0 & 21.1 \\
{\cellcolor{LightCyan}}Rank-K (Single Call) & {\cellcolor{LightCyan}}50.8 & {\cellcolor{LightCyan}}49.4 & {\cellcolor{LightCyan}}28.2 & {\cellcolor{LightCyan}}46.0 & {\cellcolor{LightCyan}}27.3 & {\cellcolor{LightCyan}}30.5 & {\cellcolor{LightCyan}}31.9 & {\cellcolor{LightCyan}}24.7 & {\cellcolor{LightCyan}}22.4 & {\cellcolor{LightCyan}}5.9 & {\cellcolor{LightCyan}}23.2 & {\cellcolor{LightCyan}}20.9 & {\cellcolor{LightCyan}}30.1 \\
\midrule
\multicolumn{14}{l}{\textit{Rerank Top 50}} \\
\midrule
RankZephyr 7B & 44.1 & 31.0 & 17.9 & 28.4 & 17.5 & 27.0 & 21.6 & 18.9 & 17.8 & 2.7 & 15.9 & 12.7 & 21.3 \\
{\cellcolor{LightCyan}}Rank-K (Single Call) & {\cellcolor{LightCyan}}48.9 & {\cellcolor{LightCyan}}44.7 & {\cellcolor{LightCyan}}29.2 & {\cellcolor{LightCyan}}47.8 & {\cellcolor{LightCyan}}30.3 & {\cellcolor{LightCyan}}30.8 & {\cellcolor{LightCyan}}35.9 & {\cellcolor{LightCyan}}22.4 & {\cellcolor{LightCyan}}24.1 & {\cellcolor{LightCyan}}7.1 & {\cellcolor{LightCyan}}26.1 & {\cellcolor{LightCyan}}26.3 & {\cellcolor{LightCyan}}31.1 \\
{\cellcolor{LightCyan}}Rank-K (Sliding Window) & {\cellcolor{LightCyan}}52.1 & {\cellcolor{LightCyan}}47.2 & {\cellcolor{LightCyan}}30.0 & {\cellcolor{LightCyan}}47.7 & {\cellcolor{LightCyan}}31.9 & {\cellcolor{LightCyan}}30.7 & {\cellcolor{LightCyan}}36.5 & {\cellcolor{LightCyan}}\textbf{25.4} & {\cellcolor{LightCyan}}27.5 & {\cellcolor{LightCyan}}6.5 & {\cellcolor{LightCyan}}27.7 & {\cellcolor{LightCyan}}26.0 & {\cellcolor{LightCyan}}32.4 \\

\midrule
\multicolumn{14}{l}{\textit{Rerank Top 100}} \\
\midrule
MonoT5-3B & 16.0 & 24.0 & 17.7 & 19.5 & 8.0 & 10.5 & 19.5 & 17.2 & 29.2 & 7.1 & 20.3 & 12.0 & 16.8 \\
RankLLaMA-7B & 17.5 & 15.5 & 13.1 & 13.6 & 17.9 & 6.9 & 16.9 & 8.4 & \textbf{46.8} & 2.2 & 4.5 & 3.5 & 13.9 \\
RankLLaMA-13B  & 21.6 & 19.1 & 16.3 & 14.0 & 15.7 & 7.7 & 18.5 & 8.8 & 31.1 & 1.7 & 4.4 & 4.9 & 13.7 \\
RankZephyr 7B & 44.1 & 31.0 & 17.9 & 28.4 & 17.5 & 27.0 & 21.6 & 18.9 & 17.8 & 2.7 & 15.9 & 12.7 & 21.3 \\
Rank1-32B & 49.7 & 35.8 & 22.0 & 37.5 & 22.5 & 21.7 & 35.0 & 18.8 & 32.5 & \textbf{10.8} & 22.9 & \textbf{43.7} & 29.4 \\

{\cellcolor{LightCyan}}Rank-K (Sliding Window) & {\cellcolor{LightCyan}}\textbf{50.4} & {\cellcolor{LightCyan}}\textbf{46.2} & {\cellcolor{LightCyan}}\textbf{30.6} & {\cellcolor{LightCyan}}\textbf{46.7} & {\cellcolor{LightCyan}}\textbf{32.4} & {\cellcolor{LightCyan}}\textbf{33.0} & {\cellcolor{LightCyan}}\textbf{41.2} & {\cellcolor{LightCyan}}24.0 & {\cellcolor{LightCyan}}32.2 & {\cellcolor{LightCyan}}7.6 & {\cellcolor{LightCyan}}\textbf{28.3} & {\cellcolor{LightCyan}}26.6 & {\cellcolor{LightCyan}}\textbf{33.3} \\

\bottomrule
\end{tabular}
}
\caption{nDCG@10 on BRIGHT. BM25 scores are taken from the official BRIGHT paper. All models rerank from the BM25 on GPT-4o CoT top 100 documents, but are not given the GPT-4o CoT. Note that RankZephyr produced exactly the same results when reranking the top 50 and 100. }
\label{tab:bright}
\end{table*}

\begin{table*}[t!]
\centering
\resizebox{\textwidth}{!}{
\begin{tabular}{l|rrrrrrr|rr|rrr|r}
\toprule
& \multicolumn{7}{c|}{StackExchange} & \multicolumn{2}{c|}{Coding} & \multicolumn{3}{c|}{Theorem-based} & \multirow{2}{*}{\centering Avg.}\\
\cmidrule(r){2-8} \cmidrule(r){9-10} \cmidrule(r){11-13}
& Bio. & Earth. & Econ. & Psy. & Rob. & Stack. & Sus. & Leet. & Pony & AoPS & TQ. & TT. \\
\midrule
ReasonIR on GPT-4o CoT &  43.5 & 43.0 & 32.7 & 38.9 & 20.8 & \textbf{30.9} & 27.3 & 31.5 & 19.7 & 7.4 & 35.0 & \textbf{36.6} & 30.6 \\
\midrule
\cellcolor{LightCyan}Rank-K (Sliding Window) & \cellcolor{LightCyan}48.4 & \cellcolor{LightCyan}38.0 & \cellcolor{LightCyan}31.5 & \cellcolor{LightCyan}\textbf{45.9} & \cellcolor{LightCyan}24.3 & \cellcolor{LightCyan}27.4 & \cellcolor{LightCyan}\textbf{40.2} & \cellcolor{LightCyan}26.2 & \cellcolor{LightCyan}\textbf{27.1} & \cellcolor{LightCyan}8.2 & \cellcolor{LightCyan}\textbf{38.3} & \cellcolor{LightCyan}27.3 & \cellcolor{LightCyan}31.9 \\

\cellcolor{LightCyan}$+$ ReasonIR with RRF & \cellcolor{LightCyan}\textbf{50.6} & \cellcolor{LightCyan}\textbf{44.2} & \cellcolor{LightCyan}\textbf{35.3} & \cellcolor{LightCyan}45.4 & \cellcolor{LightCyan}\textbf{24.7} & \cellcolor{LightCyan}30.7 & \cellcolor{LightCyan}38.4 & \cellcolor{LightCyan}\textbf{32.7} & \cellcolor{LightCyan}24.2 & \cellcolor{LightCyan}\textbf{9.0} & \cellcolor{LightCyan}\textbf{38.3} & \cellcolor{LightCyan}28.1 & \cellcolor{LightCyan}\textbf{33.5} \\

\bottomrule
\end{tabular}
}
\caption{nDCG@10 on BRIGHT when reranking the top 100 documents from ReasonIR without custom prompts for different subsets.}
\label{tab:bright-on-reasonir}
\end{table*}

We report the effectiveness of our models on BRIGHT in Table~\ref{tab:bright}.
When reranking only the top 20 documents, we observe a 42.7\% (21.1 to 30.1) improvement over RankZephyr when macro-averaging over all 12 subsets.
The gap becomes larger when we rerank the top 50 and 100 documents in a single call compared to the sliding window approach that RankZephyr uses underneath, which highlights the efficiency and test-time scalability of Rank-K. 
Interestingly, while adopting the sliding window approach further improves the effectiveness with a high efficiency cost, 
reranking 50 documents in a single Rank-K call, which is beyond what the model was trained for, still provides a decent effectiveness improvement without needing to wait to process each sequential window. 

When reranking the top 100 documents, which is the setup that other reranking works usually use, we improve 13.3\% over the recent Rank1-32B model that also leverages the reasoning LM. 
Specifically, in Rank1, since the relevance definition of Pony, AoPS, TheoremQA-Questions (TQ), and TheoremQA-Theorems (TT) are less common than the typical ad hoc retrieval scenario, many previous works \citet{shao2025reasonir,weller2025rank1} adopted separate prompts for them. 
While using separate prompts provides evidence for adaptivity and flexibility through prompting, we argue that it is less robust to general ad hoc queries, as we usually do not assume any query distribution.
With Rank-K we always use the same prompt in training and evaluation, highlighting the robustness of our approach.

Additionally, we also report results on reranking top-ranked documents from ReasonIR, a concurrent dense retrieval model that also focuses on reasoning retrieval tasks. 
Similar to results on SPLADE-v3, Rank-K is also capable of improving results from strong initial retrieval results. 
Notably, the effectiveness improves even more when fusing the reranked and initial rank lists with reciprocal rank fusion, which does not require any additional computation since the initial rank list always exists.
Note that ReasonIR also uses subset-specific prompts where Rank-K does not, which explains some degradation after reranking, such as TheoremQA-Theorems.

\begin{table}[]
\centering

\begin{tabular}{l|cc|cc|cc}
\toprule
 & \multicolumn{2}{c|}{Persian} & \multicolumn{2}{c|}{Russian} & \multicolumn{2}{c}{Chinese} \\
Reranker   & Translated & Native & Translated & Native & Translated & Native \\
\midrule
        -- & 0.386 &    -- & 0.337 &    -- & 0.375 &    -- \\
\midrule
RankZephyr & 0.291 & 0.369 & 0.279 & 0.238 & 0.274 & 0.367 \\
R1 Teacher & 0.482 & 0.474 & 0.474 & 0.468 & 0.470 & 0.470 \\
\midrule
\cellcolor{LightCyan}Rank-K & {\cellcolor{LightCyan}}0.440 & {\cellcolor{LightCyan}}0.441 & {\cellcolor{LightCyan}}0.434 & {\cellcolor{LightCyan}}0.439 & {\cellcolor{LightCyan}}0.447 & {\cellcolor{LightCyan}}0.449 \\
\bottomrule
\end{tabular}
\caption{nDCG@10 on reranking BM25 evaluated on NeuCLIR with zero-Shot language transfer to cross-language IR. Our contribution is highlighted in blue.}\label{tab:zs-to-other-langs}
\end{table}

\subsection{Zero-shot Language Transfer to Cross-Language IR}

Since the base model of Rank-K, QwQ 32B, is a multilingual model, we experiment with transferring our Rank-K model to a cross-language reranking task with English queries and documents in other languages using the native document in the NeuCLIR collection (Persian, Russian, and Chinese documents). 
The results are summarized in Table~\ref{tab:zs-to-other-langs}. 

Despite only being fine-tuned for the reranking task with English data, Rank-K can effectively rerank non-English passages based on English queries. The differences between reranking the translated and native documents are negligible and likely to be the random fluctuations. 
Surprisingly, RankZephyr, where its base model Mistral-7B is also multilingual, reranking the native documents is sometimes more effective than reranking the translation, with Chinese demonstrating the largest difference.
Since the translation may drift away from its original information in the native document, Rank-K (as well as its teacher) is more capable of analyzing the content and overcoming the upstream translation error during reasoning and reranking, again highlighting the robustness of the approach.

\subsection{Ranking Progression during Reasoning}

\begin{figure}
    \centering
    \includegraphics[width=\linewidth]{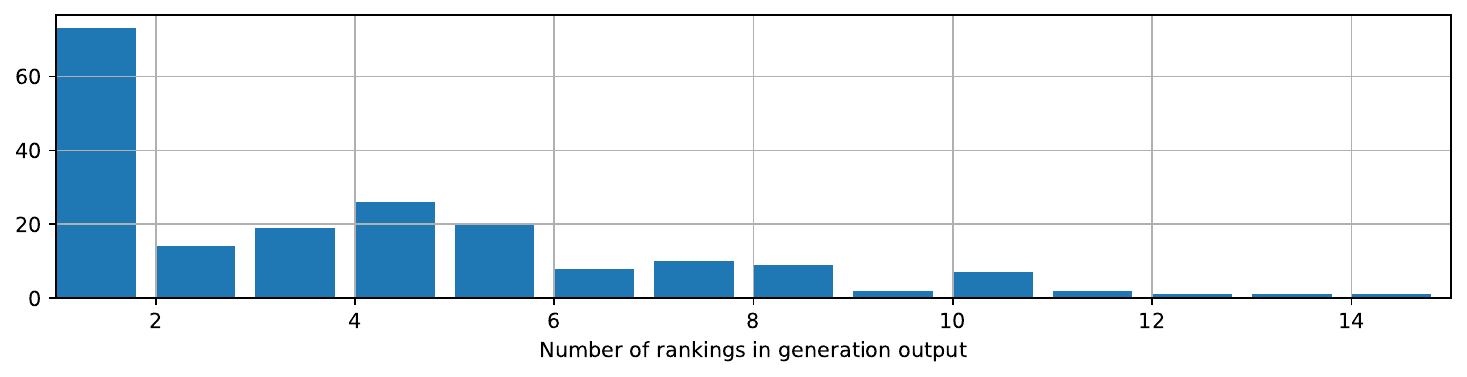}
    \caption{Histogram of the number of ranking Rank-K generates in reranking for each query on TREC DL 2019 and 2020. The count includes the intermediate partial rankings and the final full ranking. We see that Rank-K generates a non-uniform distribution of rankings.}
    \label{fig:n-ranking-hist}
\end{figure}

During the reasoning process of Rank-K, passages are categorized into buckets, and partial orderings are developed. As is illustrated in Figure~\ref{fig:n-ranking-hist}, Rank-K generates a single ranking for 38\% of the queries. This means that for almost two-thirds of queries, Rank-K needs to produce multiple rankings before assembling the final full ranking. In a handful of instances, more than ten partial rankings are created before Rank-K assembles the final ranking. Figure~\ref{example-multi-ranking} is an example of the type of reasoning Rank-K does when producing multiple partial rankings. 
The process involves breaking ties and considering the differences between passages that are similar at a high level but can still be distinguished. 
This example contains both ranking in the final format as well as partial rankings embedded in sentences without the correct output formats, which still helps improve the final ranking. 
There are cases where the model repeatedly summarizes passages during reasoning as a reminder for the content of the passage being compared, which led to longer generations than desired. 
While such generations are certainly computationally expensive, incremental reranking over reasoning provides a dynamic scaling capacity at test time to tackle a more challenging reranking problem.
We leave reducing the length of the reasoning traces for future work.

\begin{figure*}
\noindent\fbox{%
    \parbox{.98\textwidth}{%
\footnotesize
{\tt
\{...\}
So the order would be:

[13] > [14] > [19] > [3] = [6] = [7] = [8] = [9] = [15] > [1] = [2] = [4] = [5] = [20] > [10] = [11] = [12] = [16] = [17] = [18]

But need to check if any other passages have more details. For example, passage 19 also mentions season 5 locations and safety, which might make it more detailed than 14. But 14 gives a specific example with a reader's experience. Both are very relevant. So maybe [13] first, then [19] and [14] next. But [14] mentions specific streets, which is more precise. So maybe [13] > [14] > [19].

Then the group of 3,6,7,8,9,15.

Then the general ones.

So the final ranking would be:

[13] > [14] > [19] > [3] > [6] > [7] > [8] > [9] > [15] > [1] > [2] > [4] > [5] > [20] > [10] > [11] > [12] > [16] > [17] > [18]

But the user allows ties. So grouping similar ones:

[13] > [14] = [19] > [3] = [6] = [7] = [8] = [9] = [15] > [1] = [2] = [4] = [5] = [20] > [10] = [11] = [12] = [16] = [17] = [18]

But maybe [14] and [19] are equally relevant. Similarly, the group 3,6,7,8,9,15 are all about current filming in Chicago, so they can be tied. The general ones can be tied as well.

So the final order would be:

[13] > [14] = [19] > [3] = [6] = [7] = [8] = [9] = [15] > [1] = [2] = [4] = [5] = [20] > [10] = [11] = [12] = [16] = [17] = [18]

But the user wants the output as a list with numbers in order. So the correct format would be:

[13] > [14] = [19] > [3] = [6] = [7] = [8] = [9] = [15] > [1] = [2] = [4] = [5] = [20] > [10] = [11] = [12] = [16] = [17] = [18]

But wait, the user's example output was [3] > [2] > [4] = [1] > [5]. So they include all passages, but in this case
\{...\}
}
}}

\caption{A partial example thinking process produced by Rank-K. Passage summaries and self-reflection are omitted for presentation.  }
\label{example-multi-ranking}
\end{figure*}
\section{Related Work}

\subsection{Reranking Methods}

Reordering documents with a generative language model has shown remarkable success in ad hoc retrieval, especially when utilizing large instruction-tuned models like Llama and GPT-4~\citep{sun2023rankgpt,zhang-etal-2024-two,ma2024fine}.
These reranking techniques can be divided into pointwise~\citep{zhuang-etal-2023-open,ma2024fine}, pairwise~\citep{qin2024pairwise}, and listwise~\citep{sun2023rankgpt,ma2023zero,zhang-etal-2024-two} approaches. 
Pointwise methods assign a score to each query-document pair, which is then used to rank the candidates. Implementation of pointwise methods can involve sequence pair classification~\citep{monobert},
query probability generation (i.e., generating the probability of the query given the doc)~\citep{sachan2022improving},
single token generation~\citep{monot5}, etc. 

Evaluating multiple documents based on a query offers a more straightforward ranking method compared to generating independent scores, leading to a more effective ranking at a higher computational costs~\citep{pradeep2021duot5}. 
Pairwise comparison, a specific form of listwise ranking, identifies which of two documents should be ranked higher given the query. This can be achieved using either  encoder~\citep{nogueira2019duobert} or decoder~\citep{pradeep2021duot5} models.
However, pairwise ranking is significantly more computationally intensive than pointwise ranking, especially since the assumption of transitivity may not always be valid. This has led to efforts to approximate the complete ranking by reducing or sampling comparison pairs~\citep{mikhailiuk2021active, gienapp2022sparse, qin2024pairwise}. 

With the expanding context window of generative models, it becomes feasible to reduce the number of model inference calls by comparing more documents simultaneously, utilizing listwise approaches~\citep{ma2023zero, sun2023rankgpt}. 
Instruction-tuned large language models (LLMs) like GPT-4, Gemini, and Llama can generate a ranked list of document identifiers specified in the prompt~\citep{ma2023zero, sun2023rankgpt}. 
Alternatively, with appropriate training, the relative likelihood of generating each document's identifier can be interpreted as a score within the input batch or list~\citep{reddy2024first}.
This method builds upon earlier work with MonoT5~\citep{monot5} and DuoT5~\citep{pradeep2021duot5} work. 
Additionally, distilling larger models into smaller ones may reduce the compute time for each call, thereby reducing latency~\citep{tamber2023lit5, pradeep2023rankzephyr}.

One of the key aspects of listwise approaches is how to accomplish a sort that expands beyond the context of the window that the LLM can effectively order. Setwise bubble sort has been the dominant approach to sorting since \cite{sun2023rankgpt}. 
While a single pass of bubble sort has obvious limitations, 
a few iterations of bubble sort are sufficient to push non-relevant documents outside the scoring range and bubble-up relevant documents for most metrics~\citep{pradeep2023rankvicuna}. 
That said, there are other Setwise sorting algorithms, such as Setwise Heap sort, which have different trade-offs in terms of efficiency and effectiveness~\citep{zhuang2024setwise}.

\subsection{Reasoning Language Models}
OpenAI introduced reasoning language models with their o1 model \citep{jaech2024openai}, which demonstrated significantly enhanced performance on tasks requiring reasoning, such as mathematics, logic, and programming. Following their release, other organizations have developed similar models, including Google's Gemini Flash Thinking\footnote{\url{https://deepmind.google/technologies/gemini/flash-thinking/}} and Deepseek's R1~\citep{deepseekai2025deepseekr1}. There has been a surge of activity in the open-source community, both before and after the release of o1, focusing on reasoning language models and systems. This work has addressed areas such as reproduction \citep{snell2024scaling,muennighoff2025s1}, calibration and confidence \citep{jurayj2025your}, agentic capabilities with explainable reasoning traces  \citep{weir2022nellie,weir2024enhancing}, and more. We anticipate that this line of research will continue, and ongoing collaboration is likely to enhance these models' performance in retrieval tasks as well.

Recent works also explore training reranking explicitly for reasoning tasks besides general ad hoc retrieval queries~\citep{samarinas2025distillation, shao2025reasonir, zhuang2025rankr1, weller2025rank1}, which are primarily evaluated with BRIGHT~\citep{bright}. 

\section{Conclusion}

In this work, we introduced Rank-K, a test-time listwise reranking model fine-tuned from QwQ-32B. 
Evaluating on TREC Deep Learning and NeuCLIR collections, we showed that Rank-K is more effective than the previous state-of-the-art listwise reranker RankZephyr. 
Since Rank-K is naturally multilingual despite only fine-tuning with reranking objectives in English, it is also effective in reranking queries and passages in different languages.

\bibliography{colm2025_conference}
\bibliographystyle{colm2025_conference}

\end{document}